\documentclass[structabstract]{aa}  
\usepackage{graphicx}
\usepackage{txfonts}
\usepackage{natbib}
\bibpunct{(}{)}{;}{a}{}{,}
\usepackage{amssymb}
\usepackage{xcolor}
\usepackage{lipsum}
\usepackage{textcomp}
\usepackage{multirow}
\usepackage[colorlinks=true,linkcolor=blue,citecolor=blue,
bookmarks=true,bookmarksopen=true,bookmarksnumbered=true,draft=false]{hyperref}
\usepackage{enumerate}
\usepackage{url}
\usepackage{breakurl}
\usepackage{lscape}

\newcommand{\msun}{$M_{\sun}$}
\newcommand{\hour}{$^{\mathrm{h}}$}
\newcommand{\minute}{$^{\mathrm{m}}$}
\newcommand{\second}{$^{\mathrm{s}}$}
\newcommand{\snr}{SNR G15.9+0.2\xspace}

\newcommand{\xmm}{XMM-{\it Newton}\xspace}
\newcommand{\chandra}{{\it Chandra}\xspace}

\newcommand{\suzaku}{{\it Suzaku}}

\newcommand{\prelim}[1]{\textcolor{red}{PRELIM: #1}}

\newcommand{\revision}[1]{\normalfont{#1}} 
\newcommand{\revisiontwo}[1]{\normalfont{#1}} 

\defcitealias{2006ApJ...652L..45R}{R06}
\defcitealias{2014ApJ...785L..27Y}{Y14}

\begin{document}

   \title{Fe K and ejecta emission in SNR~G15.9+0.2 with XMM-\textit{Newton}
}

   \subtitle{}


	\author{Pierre Maggi   
	\and Fabio Acero   
	}

   \institute{Laboratoire AIM, IRFU/Service d'Astrophysique - CEA/DRF - CNRS - 
Universit\'e Paris Diderot, Bat. 709, CEA-Saclay, 91191 Gif-sur-Yvette Cedex, 
France\\
\email{pierre.maggi@cea.fr}
    }

   \date{Received 25 July 2016\,/\,Accepted 1 November 2016}

  \abstract
{}
{We present a study of the Galactic supernova remnant SNR G15.9+0.2 
with archival \xmm\ observations.}
{EPIC data are used to investigate the morphological and spectral 
properties of the remnant, searching in particular for supernova ejecta and Fe 
K line emission. By comparing the SNR's X-ray absorption column density with the
atomic and molecular gas distribution along the line of sight, we attempt to
constrain the distance to the SNR.}
{Prominent line features reveal the presence of ejecta. Abundance ratios of Mg, 
Si, S, Ar, and Ca strongly suggest that the progenitor of SNR G15.9+0.2 was a 
massive star with a main sequence mass likely in the range 20--25~\msun, 
strengthening the physical association with a candidate central compact object 
detected with Chandra. Using EPIC's collective power, Fe~K line emission from 
SNR G15.9+0.2 is detected for the first time. We measure the line properties and 
find evidence for spatial variations. We discuss how the source fits within the 
sample of SNRs with detected Fe~K emission and find that it is the core-collapse 
SNR with the lowest Fe~K centroid energy. We also present some caveats 
regarding the use of Fe~K line centroid energy as a typing tool for SNRs. Only a 
lower limit of 5~kpc is placed on the distance to SNR G15.9+0.2, constraining 
its age to $t_{SNR} \gtrsim 2$~kyr.}
{}

   \maketitle

\section{Introduction}
\label{introduction}

\object{SNR G15.9+0.2} was discovered as a supernova remnant (SNR) at radio 
wavelengths in Molonglo-Parkes observations by 
\citet{1973Natur.246...28C,1975AuJPA..37....1C}. Higher spatial resolution radio 
observations with VLA revealed the elongated shell-like structure of the SNR 
with a bright enhancement on the eastern border \citep{1996AJ....111.1304D}. 
The source is relatively bright in X-rays, as found in \chandra\ data by 
\citet[][hereafter \citetalias{2006ApJ...652L..45R}]{2006ApJ...652L..45R}. Its 
X-ray \mbox{morphology} favours a core-collapse (CC) origin according to 
\citet{2009ApJ...706L.106L,2011ApJ...732..114L}. The SNR was detected in 
infrared with \emph{Spitzer}/MIPS by \citet{2011AJ....142...47P}, who measured a 
dust temperature and mass of 60~K and $8.1\times 10^{-2}$~\msun, respectively.

Although 300 SNRs are now known in the Milky Way, \snr is an interesting target 
for an in-depth analysis for several reasons. First, it is likely young, $\sim 
10^3$~yr according to \citetalias{2006ApJ...652L..45R}, as suggested by its 
small angular diameter and \chandra\ X-ray analysis. This could add \snr to the 
small sample (about 15) of SNRs confirmed to be $\lesssim 2000$~yr old 
\revision{\citep[using the 
compilation\,\footnote{\url{http://www.physics.umanitoba.ca/snr/SNRcat/}} 
of][]{2012AdSpR..49.1313F}} and thus reduces the discrepancy between the 
observed and expected numbers of young SNRs. Second, a compact source, 
identified as \object{CXOU~J181852.0$-$150213}, was discovered in the central 
region of the SNR \citepalias{2006ApJ...652L..45R} with X-ray properties typical 
of central compact objects \citep[CCOs, e.g.][]{2004IAUS..218..239P}. 
\revision{Recently, deeper \chandra\ observations allowed 
\citet{2016A&A...592L..12K} to confirm that the central object is indeed a young 
cooling low-magnetized neutron star.} This adds \snr to the short list of seven 
SNRs hosting a confirmed CCO and as many hosting bona fide candidates 
\citep{2005ApJ...618..733L,2006Sci...313..814D,2008AIPC..983..311D, 
2010ApJ...709..436H,2010A&A...522A..50C,2012Ap&SS.337..573S}.
Third, its young age and plasma conditions should produce detectable Fe~K 
emission. This feature is a blend of lines from various Fe ions, with a centroid 
energy at $\sim 6.4$~keV for ionisation states lower than Fe$^{17+}$, and then 
progressing to $\sim 6.7$~keV for ion charges up to 24. The ionic fraction of 
iron, and thus the centroid energy of the Fe~K line, is chiefly governed by the 
ionisation timescale $n_e t$, where $n_e$ is the electron density and $t$ the 
time since the plasma was shocked. After analysing all SNRs observed with 
\suzaku, \citet[][hereafter 
\citetalias{2014ApJ...785L..27Y}]{2014ApJ...785L..27Y} detected Fe~K emission 
from 16 Galactic and seven LMC SNRs. They showed that the Fe~K centroid energy 
discriminates type Ia and CC SNRs, with the former consistently in the lower end 
($\sim 6.4$~keV) of the ionisation range, and CC SNRs closer to 6.7~keV. They 
attributed this to CC SNRs expanding into higher density environments produced 
by the massive star prior to the explosion, while the ambient medium around type 
Ia SNRs is hardly modified by the SN progenitors. If Fe~K emission from \snr\ 
were detected, we could obtain independent clues to the origin of the remnant.

\snr was serendipitously observed in \xmm\ pointings of a neighbouring target. 
Thanks to its superior collective power, there are several points where results 
from \citetalias{2006ApJ...652L..45R} can be elaborated upon and improved. In 
particular, the effective area of EPIC-pn is five times higher than \chandra's 
ACIS at 6.4~keV, making it a prime instrument to search for faint Fe~K emission 
yet undetected. The \chandra image revealed an incomplete shell in X-rays, 
broken in the north-west quadrant, while radio emission is still found there. 
With \xmm we can look for and characterise X-ray emission to much fainter 
limits. Furthermore, the SNR is bright and extended enough ($\sim 5.5$\arcmin\ 
across) so that even with the modest angular resolution offered by \xmm, we 
can measure abundances and plasma conditions in various regions around the 
remnant. The derived abundance ratios can constrain the type of progenitor and 
provide additional hints to the origin of the SNR. Finally, using X-ray spectral 
analysis and other tracers (namely atomic and molecular gas), we can reassess 
the age of the SNR and the measurements of its distance.

This work is organised as follows: We first present the observations 
and data reduction in Sect.\,\ref{observations}. Next, we describe in 
Sect.\,\ref{results} the results of the morphological and spectral analyses. In 
Sect.\,\ref{discussion}, we discuss the type of progenitor, age, and distance 
to \snr, as well as the evolution of Fe~K lines in SNRs in general. We 
summarise our findings in Sect.\,\ref{summary}.

\begin{table}[t]
\caption{Details of the X-ray observations}
\label{table_obs}
\centering
\begin{tabular}{c c c c}
\hline\hline
\noalign{\smallskip}
ObsID & Date & Total\,/\,filtered $t_{\mathrm{exp}}$
\tablefootmark{a} & Mode \tablefootmark{b}\\
\noalign{\smallskip}
\hline
\noalign{\smallskip}
0406450201 & 2006 Apr 6 & 43\,/\,33 & SW \\
0505240101 & 2008 Mar 31 & 93\,/\,47 & FF \\
\noalign{\smallskip}
\hline
\end{tabular}
\tablefoot{
\tablefoottext{a}{Performed duration (total) and useful (filtered) exposure
times in ks, after removal of high background intervals.}
\tablefoottext{b}{FF: full frame; SW: small window.}
}
\end{table}

\section{X-ray observations and data reduction}
\label{observations}

\snr was in the field of view (FoV) of two \xmm observations targeting PSR
J1819$-$1458, the first X-ray counterpart to a Rotating RAdio Transients
\citep[RRATs,][]{2006ApJ...639L..71R,2007ApJ...670.1307M,2013ApJ...776..104M}.
The SNR is located at off-axis angles ranging from 8\arcmin\ to 12\arcmin. Only
MOS data are available for the first observation, which was performed with EPIC
in small window (SW) mode (the outer MOS CCDs are active even in SW mode). We
discarded two additional observations that were very short ($\lesssim 5$~ks).
Details of the observations used in this paper are listed in
Table~\ref{table_obs}.

We used the XMM SAS\,\footnote{Science Analysis Software,
\url{http://xmm.esac.esa.int/sas/}} version 14.0.0 for the data reduction. We
applied a threshold of 8 and 2.5 cts~ks$^{-1}$\,arcmin$^{-2}$ on pn and MOS
light curves in the 7--15 keV energy band to screen out periods of high
background activity. This resulted in useful exposure times of 47~ks and 33~ks.

We created two sets of images and exposure maps in various energy bands from the 
filtered event lists: A ``broad band'' set covering 0.2~keV to 12~keV in five 
bands as given in \citet[their Table~3]{2009A&A...493..339W}, and an ``SNR'' set 
described in Sect.\,\ref{results_morphology}. Single and double-pixel events ($0 
\leq$ \texttt{PATTERN} $\leq 4$) were extracted from the pn detector, while all 
valid events from the MOS detectors were selected ($0 \leq$ \texttt{PATTERN} 
$\leq 12$). Masks were applied to filter out bad pixels and columns. The SAS 
task \texttt{edetectchain} is applied simultaneously to the five images of the 
broad band set to identify X-ray (point) sources in each observations. The 
detection lists were primarily used to exclude unrelated point sources from the 
spectrum extraction regions.


\begin{figure}[t]
    \centering
\includegraphics[bb= 48 152 558 640, clip,width=\hsize]
{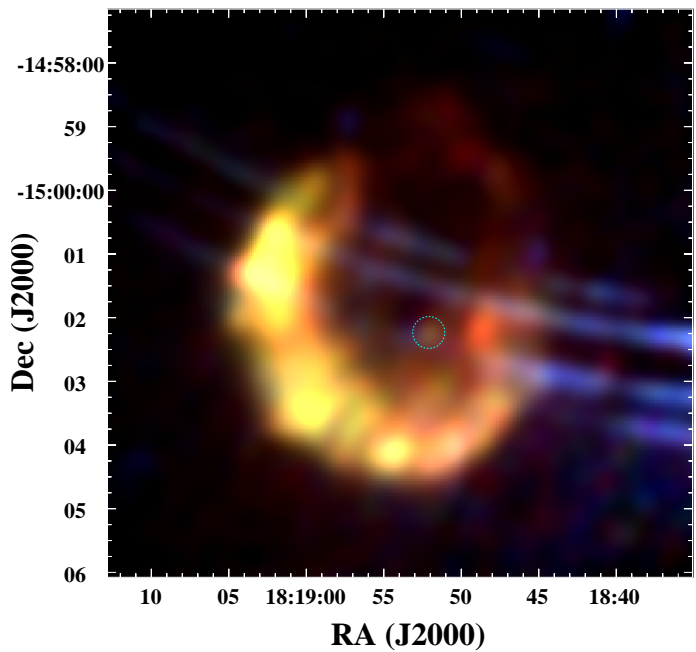}
\caption{X-ray colour image of 
\snr, combining pn and MOS data. The red, green, 
and blue components are soft, medium, and hard X-rays, as defined in 
Sect.\,\ref{results_morphology}. The position of the CCO is circled in cyan. The 
blue arcs are straylight contamination (see Sect.\,\ref{results_spectral}).}
\label{fig_xray_image}
\end{figure}

\section{Results}
\label{results}

\subsection{Morphology}
\label{results_morphology}
To study the morphology of \snr, we used three energy bands tailored to its 
spectrum \citepalias{2006ApJ...652L..45R}. A soft band from 0.9
to 2.1 keV which includes strong lines from magnesium and silicon; a medium band
from 2.1 to 3.25 keV which comprises sulphur and argon lines; and a hard band
(3.25--7.2 keV) which includes the high-energy part of the continuum and
possibly emission from calcium and Fe~K lines. Owing to high absorption, 
there is little emission below 0.8~keV, while the instrumental background 
dominates above 7.2 keV.

We used filter wheel closed (FWC) 
data\,\footnote{\url{http://www.cosmos.esa.int/web/xmm-newton/filter-closed}}, 
obtained with the detector shielded from 
astrophysical background, to subtract the detector background. The contribution 
of detector background in each observation was estimated from the count rate in 
the corners of the images, since they are not exposed to the sky. We then 
subtracted appropriately scaled FWC data from the raw images. The detector 
background-subtracted images from the two observations were merged together. 
They were then adaptively smoothed: the sizes of Gaussian kernels were chosen at 
each position to reach a signal-to-noise ratio of five, with a minimum full 
width at half maximum (FWHM) of 20\arcsec. In each band, we co-added the 
smoothed images from pn and MOS, and divided the resulting image by the 
corresponding vignetted exposure map\,\footnote{To produce a combined exposure 
map, we weighted the MOS exposure maps with a factor of 0.4 relative to pn, 
accounting for the lower effective area}.

The composite X-ray image of \snr\ is shown in Fig.\,\ref{fig_xray_image}. The 
annular stripes at high energy (in blue) are straylight emission (see 
Sect.\,\ref{results_spectral}) unrelated to the SNR. The position of the CCO is 
marked. A faint point source is detected 2\arcsec\ from CXOU~J181852.0$-$150213, 
i.e. well within the typical \xmm statistical and systematical position 
uncertainties. The flux and hardness ratios are consistent with those measured 
with \chandra \citepalias{2006ApJ...652L..45R} so we are confident that we are 
indeed detecting the CCO. However, there are not enough counts ($\lesssim 200$) 
to improve on results from \citetalias{2006ApJ...652L..45R} or 
\citet{2016A&A...592L..12K}, especially with the lower angular resolution of 
\xmm.

\snr\ exhibits a well-defined shell morphology. The eastern and south-western 
edges are particularly bright, and all the SNR emission above $\sim 3.5$~keV is 
concentrated in these regions. In contrast, the north-western quadrant of the 
shell is much fainter than the rest of the SNR, and only X-rays below 3~keV are 
detected in this region. The shell's morphology is a slightly elongated ellipse, 
with major and minor axes of 6.2\arcmin\ and 5.2\arcmin, respectively. That 
translates into a linear size of (9.0~pc~$\times$~7.6~pc) ($D$\,/\,5~kpc). The 
major axis has a position angle 150\textdegree\ eastwards of north. The centre 
of the ellipse is at RA (J2000) = 18\hour\,18\minute\,53.8\second, DEC = 
$-$15\degr\,01\minute\,38\second. The CCO is thus offset 44\arcsec\ from the 
visual centre. 

\begin{figure}[t]
    \centering
\includegraphics[width=\hsize]{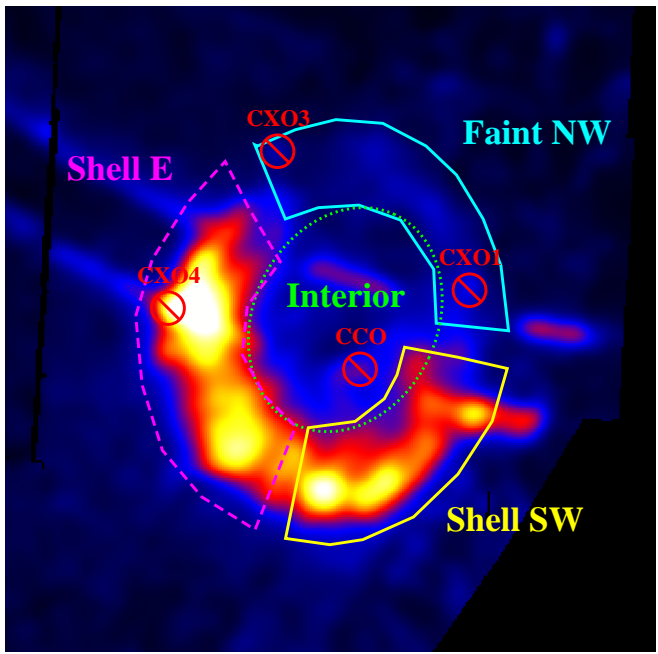}
\caption{Regions used in the spectral analysis defined on a 0.9-7.2~keV band 
image of \snr. Point sources in red were excluded. We note the changing 
straylight contamination in different regions.}
\label{fig_xray_extraction}
\end{figure}

\subsection{Spectral analysis}
\label{results_spectral}

The definition of extraction regions for spectral analysis was driven by the 
observed morphology. A spatially integrated spectra was extracted from the 
ellipse described previously. To look for spectral variations, we separated the 
bright regions of the shell in two subregions, ``Shell E'' and ``Shell SW'' (for 
east and south-west, respectively). These are shown over an X-ray image in 
Fig.\,\ref{fig_xray_extraction}. A third region covers the fainter 
north-western part of the shell (``faint NW''). We also extracted a spectrum 
from the interior of the SNR. 

Point sources detected with \chandra\footnote{From the list given in 
the \chandra\ SNR catalogue, maintained by Fred Seward\,: 
\url{http://hea-www.cfa.harvard.edu/ChandraSNR/G015.9+00.2/}}, including the 
CCO, were excised from our \xmm spectral extraction regions. Most of the sources 
detected with \xmm\ in the shell region (by \texttt{edetectchain}) are emission 
knots, therefore truly originating from the SNR and were kept in the analysis. 
Background spectra were taken from a large region surrounding the SNR (not shown 
in Fig.\,\ref{fig_xray_extraction}). \revision{We excluded straylight stripes 
and point sources detected with \xmm\ from the background regions}.

Owing to the telescope vignetting and off-axis position of the SNR, the 
effective area changes across the extent of the SNR; it decreases with larger 
off-axis angles, and more so at higher energies. Before extracting spectra, we 
corrected the filtered event lists for vignetting with the SAS task 
\texttt{evigweight}. As for the images, single and double-pixel events were 
included in pn spectra, and events with \texttt{PATTERN} = 0 to 12 were selected 
in MOS spectra. With the FTOOLS task \texttt{grppha}, we rebinned all spectra to 
have a minimum of 25 counts per bin in order to allow the use of the 
$\chi^2$-statistic. Non-rebinned spectra were used with the C-statistic 
\citep{1979ApJ...228..939C} for the study of Fe~K lines 
(Sect.\,\ref{results_spectral_FeK}) because of the limited photon statistics 
above 6~keV. These spectra were extracted from non-vignetting-weighted event 
lists to retain the Poissonian nature of the errors. XSPEC  
\citep{1996ASPC..101...17A} version 12.9.0e was used for the spectral analysis. 
All uncertainties listed in this paper are given at the 90\% confidence level, 
unless otherwise stated.

For spectral analysis, we employed the method described in 
\citet{2016A&A...585A.162M}: We simultaneously fit the source and background 
spectra with the instrumental and astrophysical contributions to the background 
explicitly modelled. This is preferable to simply subtracting a background 
spectrum taken from a nearby region because of the different instrumental 
responses and background contributions from different regions and because of 
the resulting loss in the statistical quality of the source spectrum.

Spectra were extracted from FWC data at the same detector positions as the 
source and background regions, in order to capture the spatial variations of the 
instrumental background. Our model for the instrumental background takes into 
account electronic noise and particle-induced background, as described in 
\citet{2012PhDT......ppppS} and \citet{2016A&A...585A.162M}. We first fit this 
model to FWC data from all extraction regions. The best-fit models are used in 
subsequent fits (including astrophysical signal), allowing only a constant 
renormalisation factor. This avoids overloading the number of free parameters. 
Another non-X-ray background component is the soft proton contamination (SPC), 
which we modelled following the prescription of \citet{2008A&A...478..575K}. 
The SPC parameters were different for each instrument and observation (the soft 
proton flux is highly time-variable). Indeed, we found a higher SPC in the 2006 
observation.

Next, we defined a model for the astrophysical X-ray background (AXB). 
Initially, we used a single unabsorbed thermal plasma (\texttt{apec} in XSPEC) 
for local contributions (local hot bubble and/or solar wind charge-exchange 
emission) and an absorbed two-temperature plasma (\texttt{apec + apec}) for 
the remote (Galactic) component. The cosmic X-ray background is added, with 
additional absorption, to the remote component as a power law with a photon 
index $\Gamma$ fixed to 1.41 \citep{2004A&A...419..837D}. Initial trial fits to 
the background spectra showed that a second remote Galactic component was not 
needed. This usually accounts for the very hot ($kT > 5$ keV) plasma from the 
Galactic ridge X-ray emission (GRXE), which is not detected in our data. In 
particular, we do not see the strong Fe~K line at 6.7~keV associated with the 
GRXE \citep{1986PASJ...38..121K}. \revisiontwo{We place a $3 \sigma$ upper 
limit of $2.8 \times 10^{-8}$~ph~cm$^{-2}$~s$^{-1}$~arcmin$^{-2}$ on a Galactic 
ridge Fe~K line.} Below 1~keV, the initial model had strong 
residuals. A much better fit was achieved with the inclusion of a second 
thermal component. The final model for the AXB is
\begin{equation}
S_{\mathrm{AXB}} =
S ^1 _{\mathrm{apec}} + S ^2 _{\mathrm{apec}} + \mathrm{\texttt{phabs}} (N_H  
^1)
\left(
S ^3 _{\mathrm{apec}} + \mathrm{\texttt{phabs}} (N_H  ^2) N_{\mathrm{CXB}} 
E^{-\Gamma}
\right)
\label{eq_AXB}
\end{equation}
where $S ^i _{\mathrm{apec}}$ is the emission from an \texttt{apec} model at 
temperature $kT^i$ with normalisation $Norm ^i$. The latter is defined as 
$(10^{-14} / 4 \pi D^2) \int n_e n_H dV$, with $D$ the distance to the source 
and $n_e$ and $n_H$ the density of electrons and protons, all in the cgs system 
of units. \texttt{phabs} is the photoelectric absorption model we used in 
XSPEC, with cross-sections from \citet{1992ApJ...400..699B}. 
\revision{Abundances were set 
to those of \citet{2000ApJ...542..914W}.}

\begin{figure}[t]
    \centering
\includegraphics[bb=75 10 590 700,clip,angle=-90, 
width=\hsize]{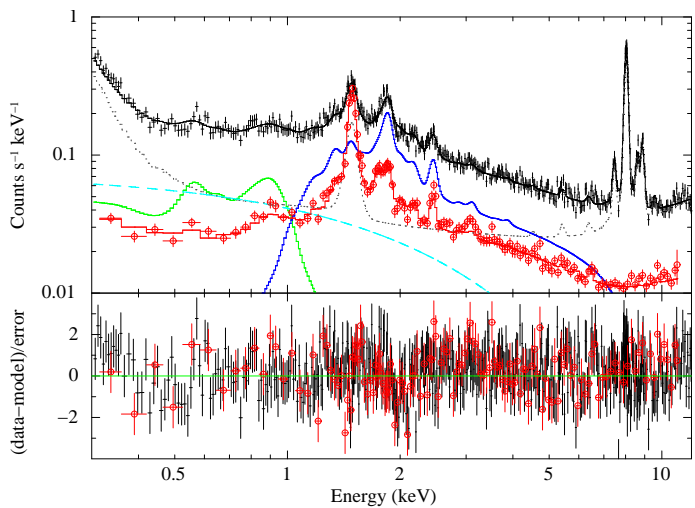}
\caption{Background spectrum of \snr. For the sake of clarity, only pn and MOS2 
data from the 2008 observation are shown (black and red points, respectively). 
Data have been rebinned for display purposes. The various components are 
instrumental background (dotted grey), local background (solid green), remote 
background and CXB (solid blue) and SPC (dashed cyan). Fit residuals are 
displayed in the bottom panel.}
\label{fig_background}
\end{figure}

\revision{The model of Eq.\,\ref{eq_AXB} reproduces fairly well the background 
spectrum (see Fig.\,\ref{fig_background}). The best-fit parameters are listed in 
Table~\ref{table_AXB}. Our CXB surface brightness is an order of magnitude 
higher than derived at high Galactic latitudes from various instruments 
\citep{2002PASJ...54..327K,2002A&A...389...93L,2004A&A...419..837D, 
2005A&A...444..381R}. This can be attributed partly to cosmic variance, and 
partly to contribution by unresolved hard X-ray sources in the Galaxy which are 
by design not in the aforementioned CXB studies. We note that 
\citet{2015ApJ...814...29K} fit the CXB on a background spectrum around RX 
J1713.7-3946, located symmetrically from \snr with respect to the Galactic 
Centre. Their value of $(1.0 - 3.2) \times 
10^{-5}$~ph~keV$^{-1}$\,cm$^{-2}$\,s$^{-1}$\,arcmin$^{-2}$ at 1 keV is fully 
consistent with ours ($1.6 \times 10^{-5}$).}

Another potentially problematic source of background is straylight. Photons from 
bright sources \emph{outside} the FoV can be singly reflected by the hyperboloid 
mirrors and reach the camera. Sources 0.3\textdegree\ to 1.4\textdegree\ 
off-axis produce ring-like structures in EPIC images\,\footnote{See 
\url{http://www.star.le.ac.uk/~amr30/BG/mjf.pdf}.}, such as seen in our images. 
Following the direction to the foci of the straylight rings from the aimpoint of 
the observations, we can firmly identify the contaminating source as GX 17+2, 
located 1.26\textdegree\ off-axis. GX~17+2 is a low-mass X-ray binary accreting 
at Eddington luminosity \citep[e.g.][and references 
therein]{2012ApJ...756...34L}. Excising straylight rings from the background 
extraction regions is straightforward. However, some single-reflection arcs also 
intersect (bright) regions of the SNR. We therefore chose to include another 
component to model the straylight emission possibly included in the spectra from 
the SNR regions. To analyse the straylight spectrum, we extracted counts from 
the ring-like structures in all the FoV, except over \snr. The spectrum of 
GX~17+2 can be represented as a combination of black body, multicolour disk, and 
power-law emission, with variable relative contributions 
\citep{2010ApJ...720..205C,2012ApJ...756...34L} depending on the state of 
GX~17+2, although a single component model is sufficient given the limited count 
statistics of the single reflections. We note in particular that we do not see 
Fe K emission from GX 17+2 \citep{2012ApJ...755...27C} in the straylight 
spectrum, which has almost no signal above 6~keV. We used an absorbed black body 
with $N_H = 3.1 \times 10^{22}$~cm$^{-2}$ and $kT=0.99$~keV. The normalisation 
is free for all exposures, as it depends on both the intrinsic variability of 
the source and on its exact location relative to the mirrors. Overall, the flux 
collected in all the arcs is $10^{-4} - 10^{-3}$ times the flux of GX~17+2, 
consistent with the straylight collecting area (\textless 0.2~\% of the on-axis 
effective area). We included this component in the background model of spectra 
in the source regions. The spectral parameters were fixed. A constant factor was 
the only free parameter allowed to vary between 0 (no contribution) and 
$f_{\mathrm{stray}}$, where $f_{\mathrm{stray}}$ is a geometrical factor, the 
fraction of the straylight arc overlapping the SNR region.

\begin{table}[t]
\caption{Best-fit parameters for the AXB model of Eq.\,\ref{eq_AXB}.}
\label{table_AXB}
\centering
\begin{tabular}{l c}
\hline\hline
\noalign{\smallskip}
Parameter & Value \\
\noalign{\smallskip}
\hline
\noalign{\smallskip}
\multicolumn{2}{c}{$\chi ^2$ / d.o.f ($\chi^2_r$) $= 3089.8/2712 \ (1.14)$} \\
\noalign{\smallskip}
\hline
\noalign{\smallskip}
\multicolumn{2}{c}{Local background} \\
$kT^1$ (keV) & $0.19 \pm 0.02$ \\
$Norm ^1$ (cm$^{-5}$) & $(1.95 \pm 0.22) \times 10^{-5}$ \\
SB$^1$ (erg cm$^{-2}$ s$^{-1}$ arcmin$^{-2}$) & $1.8 \times 10^{-15}$ \\
\noalign{\smallskip}
$kT^2$ (keV) & $0.84 \pm 0.04$ \\
$Norm ^2$ (cm$^{-5}$) & $(1.95 \pm 0.22) \times 10^{-5}$ \\
SB$^2$ (erg cm$^{-2}$ s$^{-1}$ arcmin$^{-2}$) & $2.3 \times 10^{-15}$ \\
\noalign{\smallskip}
\multicolumn{2}{c}{Remote background} \\
\noalign{\smallskip}
$N_H  ^1$ ($10^{22}$ cm$^{-2}$) & $3.14 \pm 0.10$ \\
$kT^3$ (keV) & $0.72 \pm 0.03$ \\
$Norm ^3$ (cm$^{-5}$) & $(5.05_{-0.56}^{+0.44}) \times 10^{-3}$ \\
SB$^3$ (erg cm$^{-2}$ s$^{-1}$ arcmin$^{-2}$) & $2.4 \times 10^{-14}$ \\
\noalign{\smallskip}
\multicolumn{2}{c}{Cosmic X-ray background} \\
\noalign{\smallskip}
$N_H  ^2$ ($10^{22}$ cm$^{-2}$) & $1.25_{-0.73}^{+0.93}$ \\
$\Gamma$ & 1.41 (fixed) \\
$N_{\mathrm{CXB}}$ (ph keV$^{-1}$\,cm$^{-2}$\,s$^{-1}$ at 1 keV )& $(2.84 
\pm 0.13) \times 10^{-4}$\\
SB$_{\mathrm{CXB}}$ (erg cm$^{-2}$ s$^{-1}$ arcmin$^{-2}$) & $6.5 \times 
10^{-14}$ \\
\noalign{\smallskip}
\hline
\end{tabular}
\tablefoot{For each component the surface brightness (SB) is given in the 
energy band 0.3--8~keV.
}
\end{table}

\begin{figure*}[t]
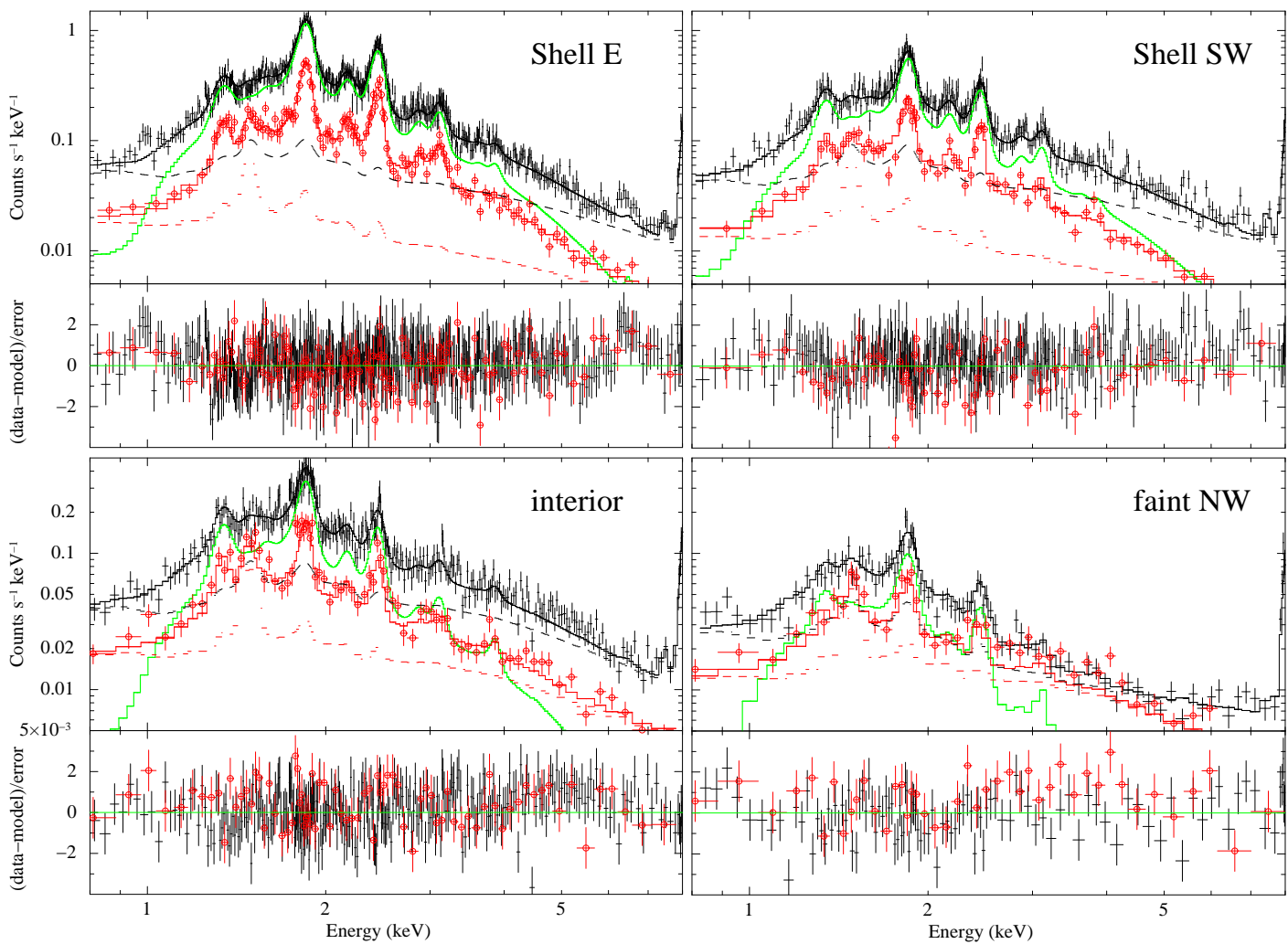

  \begin{center}
    \includegraphics[angle=-90,width=0.5286\hsize,bb = 80 10 532 700, clip]
    {final_shellE_bg_label.ps}
    \includegraphics[angle=-90,width=0.4664\hsize,bb = 80 92 532 700, clip]
    {final_shellSW_bg_label.ps}

    \includegraphics[angle=-90,width=0.5286\hsize,bb = 75 10 590 700, clip]
    {final_interior_bg_label.ps}
    \includegraphics[angle=-90,width=0.4664\hsize,bb = 75 92 590 700, clip]
    {final_faintNW_bg_label.ps}
  \end{center}
  \caption{X-ray spectra from various regions of \snr, with pn and MOS2 data 
from the 2008 observation in black and red, respectively. All background 
components are included in the dashed lines (model from 
Fig.\,\ref{fig_background}, plus straylight contamination, see 
Sect.\,\ref{results_spectral}). The best-fit fiducial model for the SNR 
emission is in green, with corresponding residuals in the bottom panels. We 
note the changing vertical scale between rows.}
  \label{fig_spectra}
\end{figure*}

\subsubsection{Spectral fits and spatial variations}
\label{results_spectral_fits}
In a first attempt, we fit single-temperature models to the spectra from each 
region, combining the background model and one or more components for the SNR 
emission. All background parameters are fixed to their best-fit values, except 
for normalisation factors for the AXB and the straylight models. We assumed 
non-equilibrium ionisation (NEI) for the SNR emission, using a plane-parallel 
shock model  \citep[\texttt{vpshock} in XSPEC, ][]{2001ApJ...548..820B}, which 
features a linear distribution of ionisation ages ($\tau = \int _0 ^t n_e dt$). 
AtomDB 3.0\,\footnote{\url{http://www.atomdb.org/index.php}} is used 
to calculate the spectrum.

We started with the ``Shell E'' spectrum from the brightest region. Initial 
fits with \texttt{vpshock} and solar abundances clearly failed, leaving strong 
residuals at line energies of main elements, in particular Si and S. With free 
Mg, Si, and S abundances, the fit improved by $\Delta \chi ^2 = -797$. Some 
residuals remained at $E \sim 3.2$~keV and 3.9~keV, where Ar and Ca emission is 
expected. With their abundances freed, the fit improved further by $\Delta \chi 
^2 = -95$ for a final $\chi ^2 $ / d.o.f $ = 2191.2/1967$. The enhanced 
abundances are a tell-tale sign of the shock-heated ejecta origin of the 
emission. In all these fits there are line-like residuals at 6.4~keV -- 6.7~keV, 
which betrays unaccounted for Fe~K emission. It cannot be improved by enhanced 
Fe abundance, because the plasma temperature, constrained by the bulk of the 
emission, is too low to produce appreciable Fe~K emission. Adding a Gaussian 
improved the fits; we analyse the Fe~K emission in detail in 
Sect.\,\ref{results_spectral_FeK}.

Because the 6.4~keV feature is very minor, we restrict in a first time the 
analysis to the \texttt{vpshock} model with free Mg, Si, S, Ar, and Ca 
abundances as it gives a good fit to the the global emission that can be applied 
in all regions of the SNR. We refer to this model as the fiducial model. The 
best-fit parameters in each region are listed in 
Table~\ref{table_spectral_results}.

\begin{table*}[t]
\caption{Best-fit parameters for the fiducial model.}
\label{table_spectral_results}
\centering
\begin{tabular}{@{\hspace{0.30cm}} l @{\hspace{0.40cm}} @{\hspace{0.40cm}} c 
@{\hspace{0.45cm}} @{\hspace{0.6cm}} c @{\hspace{0.45cm}} @{\hspace{0.50cm}} c 
@{\hspace{0.20cm}} @{\hspace{0.20cm}} c @{\hspace{0.15cm}} @{\hspace{0.15cm}} c 
@{\hspace{0.30cm}} }
\hline
\hline
\noalign{\smallskip}
Region & $N_{H}$ & $kT$ & $\tau$ & \emph{Norm} & $\chi ^2 $ / d.o.f. \\
 &($10^{22}$ cm$^{-2}$) & (keV) & ($10^{10}$ s\,cm$^{-3}$) & (cm$^{-5}$) & \\
\noalign{\smallskip}
\hline
\noalign{\smallskip}
Shell E & 5.03$_{-0.25}^{+0.21}$ & 1.32$_{-0.09}^{+0.08}$ & 
8.54$_{-0.81}^{+0.99}$ & $(1.32_{-0.14}^{+0.20}) \times 10^{-2}$ & 
2191.2/1967 (1.11) \\
\noalign{\smallskip}
Shell SW & 4.30$_{-0.27}^{+0.17}$ & 1.53$_{-0.09}^{+0.28}$ & 
7.69$_{-1.25}^{+1.18}$ & $(5.18_{-1.17}^{+0.58}) \times 10^{-3}$ & 
1980.2/1726 (1.15) \\
\noalign{\smallskip}
Interior & 4.10$_{-0.26}^{+0.12}$ & 1.46$_{-0.21}^{+0.28}$ & 
6.69$_{-1.39}^{+1.56}$ & $(3.15_{-0.74}^{+0.83}) \times 10^{-3}$ & 
1678.3/1632 (1.03) \\
\noalign{\smallskip}
Faint NW & 3.77$_{-0.35}^{+0.44}$ & 0.67$\pm 0.20$ & 96.8 $(> 34.7)$ & 
$(2.59_{-1.05}^{+3.13}) \times 10^{-3}$ & 
959.4/896 (1.07) \\
\noalign{\smallskip}
Integrated & 4.58$\pm 0.11$ & 1.44$_{-0.06}^{+0.08}$ & 7.85$_{-0.60}^{+0.79}$
&  $(2.34_{-0.19}^{+0.16}) \times 10^{-2}$ & 
3860.4/3418 (1.13) \\
\noalign{\smallskip}
\hline
\noalign{\smallskip}
& Mg & Si & S &  Ar & Ca \\
\noalign{\smallskip}
\hline
\noalign{\smallskip}
Shell E & 1.79$_{-0.25}^{+0.32}$ & 2.57 $(\pm 0.18)$ & 2.69 $(\pm 0.17)$ & 
3.44$_{-0.45}^{+0.51}$ & 2.53$_{-1.02}^{+1.22}$ \\
\noalign{\smallskip}
Shell SW & 1.79$_{-0.30}^{+0.40}$ & 2.06$_{-0.17}^{+0.30}$ & 
2.11$_{-0.20}^{+0.29}$ & 2.73$_{-1.24}^{+0.80}$ & 1.24 $(< 3.05)$ \\
\noalign{\smallskip}
Interior & 1.83$_{-0.33}^{+0.44}$ & 2.06$_{-0.24}^{+0.32}$ & 
2.03$_{-0.27}^{+0.33}$ & 2.99$_{-1.10}^{+1.27}$ & 6.68$_{-3.62}^{+4.39}$ \\
\noalign{\smallskip}
Faint NW & 1.33$_{-0.44}^{+0.60}$ & 1.66 $(\pm 0.40)$ & 2.04$_{-0.57}^{+1.08}$ 
& 4.26$_{-2.62}^{+6.71}$ & 3.49 $(< 20.2)$ \\
\noalign{\smallskip}
Integrated & 1.74$_{-0.18}^{+0.20}$ & 2.20$_{-0.10}^{+0.13}$ & 
2.29$_{-0.11}^{+0.14}$ & 3.15$_{-0.37}^{+0.42}$ & 2.49$_{-0.91}^{+0.98}$ \\
\noalign{\smallskip}
\hline
\end{tabular}
\tablefoot{The abundances are given relative to the solar values as listed in
\citet{2000ApJ...542..914W}.
}
\end{table*}

The temperature and ionisation age in the ``Shell E'' spectrum are 1.32~keV and 
$8.5 \times 10^{10}$~cm\,s$^{-3}$.  The spectrum in the ``Shell SW'' region is 
similar to its eastern counterpart, with a higher temperature (1.53~keV) but the 
same ionisation age within the uncertainties. The interior spectrum is virtually 
the same as that in the ``Shell SW'', albeit fainter, indicating that we are 
seeing (in projection) similar shock conditions as in the south-west. The 
spectrum in the ``faint NW'' region is the most different; the temperature is 
significantly lower ($kT \approx 0.7$~keV) and the ionisation age is higher 
($\tau \sim 8 \times 10^{11}$~cm\,s$^{-3}$), although poorly constrained. The 
eastern emission is the most absorbed with a column density $N_H = 5 \times 10 
^{22}$~cm$^{-2}$, while $N_H$ is lower in the interior and south-west. 
\revision{The $N_H$ in the interior ($4.10^{+0.12}_{-0.26} \times 10 
^{22}$~cm$^{-2}$) is consistent with the value fit by 
\citet{2016A&A...592L..12K} for the CCO using a carbon atmosphere model, the 
only one that yields a reasonable distance ($10^{+9}_{-5}$~kpc).} The X-ray 
emission from the north-west appears the least absorbed ($N_H \lesssim 4 \times 
10 ^{22}$~cm$^{-2}$). These absorption features are discussed in relation with 
atomic and molecular gas in Sect.\,\ref{discussion_age_distance}.

The abundances are highly elevated (in excess of 1.5--2 times solar) in every 
region except the ``faint NW'', where some abundances are consistent with solar 
when taking into account the large uncertainties. We do not find a solar value 
for Si, as did \citetalias{2006ApJ...652L..45R} in \chandra\ data fit with the 
same model. The abundances of Ar and Ca are, however, poorly constrained in 
spectra with lower statistics, in particular in the interior and north-western 
regions. Within the uncertainties, there are no clear spatial variations of the 
abundance ratios of Mg or S relative to Si. We also applied the fiducial model 
to the integrated spectrum of \snr. The resulting best-fit parameters are listed 
in Table~\ref{table_spectral_results}. They are consistent with flux-averaging 
of the parameters obtained in the four regions.

It appears from the fiducial model that a significant fraction, or the majority, 
of the emission originates from shock-heated ejecta. Therefore, we attempted a 
multicomponent model approach \citep[similar e.g. to that 
in][]{2010A&A...519A..11K}: A single $kT$ NEI component (\texttt{vpshock}) is 
used for each element or group of elements. This examines whether various 
elements are found at different temperatures or ionisation ages.
To model emission from pure ejecta, we set the abundance of the corresponding 
element(s) to $10^9$ and all the others to 0. The model thus comprises 
\emph{i)} a component for Mg; \emph{ii)} a component for Si, S, Ar, and Ca. They 
are grouped together as they are expected to be roughly co-spatial in the ejecta 
of a (core-collapse) supernova. The abundances of S, Ar, and Ca are free 
relative to Si; \emph{iii)} a component for iron, the innermost ejecta, which we 
detected through Fe~K emission; and \emph{iv)} an additional component with 
solar abundance to account for the emission from shocked ambient medium 
(hereafter solar component). All components share a common $N_H$.

This model was applied to the bright eastern and south-western spectra. Taking 
into account the seven additional free parameters, the fits with the 
multicomponent model are equally as good or slightly better than those with the 
fiducial model. Again, the $N_H$ is higher in the eastern region, confirming 
that this spatial variation is not strongly model dependent. The 
solar-abundances component dominates the continuum.
It is therefore unsurprising that a pure-ejecta model (without the 
solar-abundances component) fails to describe the data satisfactorily. The 
temperature of the solar-abundances component is (1.3\,--\,1.5)~keV without 
spatial variation in $\tau$, as with the fiducial model. The Si-group component 
has a lower temperature (0.9\,--\,1~keV) and higher $\tau$ ($2 \times 
10^{11}$~cm\,s$^{-3}$) than the solar and Mg components. Without the help of the 
absorbed iron L-shell lines below 1.2~keV, we cannot constrain well the $kT - 
\tau$ combination for Fe, an issue further complicated by the complex nature of 
the Fe~K emission, as described in the next section.


\subsubsection{Fe~K emission}
\label{results_spectral_FeK}
Here we present a separate treatment of the Fe~K emission at 6.4~keV -- 6.7~keV 
for comparison with the results of \citetalias{2014ApJ...785L..27Y}. The same 
background model as previously described is included. Data between 0.6~keV and 
4~keV are excluded, thus avoiding the prominent line emission from elements 
other than Fe. The low-energy part (from 0.3 to 0.6~keV) is used to fix the 
normalisation of the background model. The continuum emission is modelled by an 
absorbed Bremsstrahlung ($N_H$ fixed to $5 \times 10^{22}$~cm$^{-2}$), and the 
Fe~K emission by a Gaussian. There is only a significant detection of the line 
in the ``Shell E'' and ``Shell SW'' regions, and particularly in pn data. There 
is no clear signature in the MOS spectra owing to their lower effective area. 
Still, we present the results of the joint pn/MOS analysis since it helps in
constraining the continuum. No differences were found in the line properties 
when using only pn data.

\begin{figure*}[t]
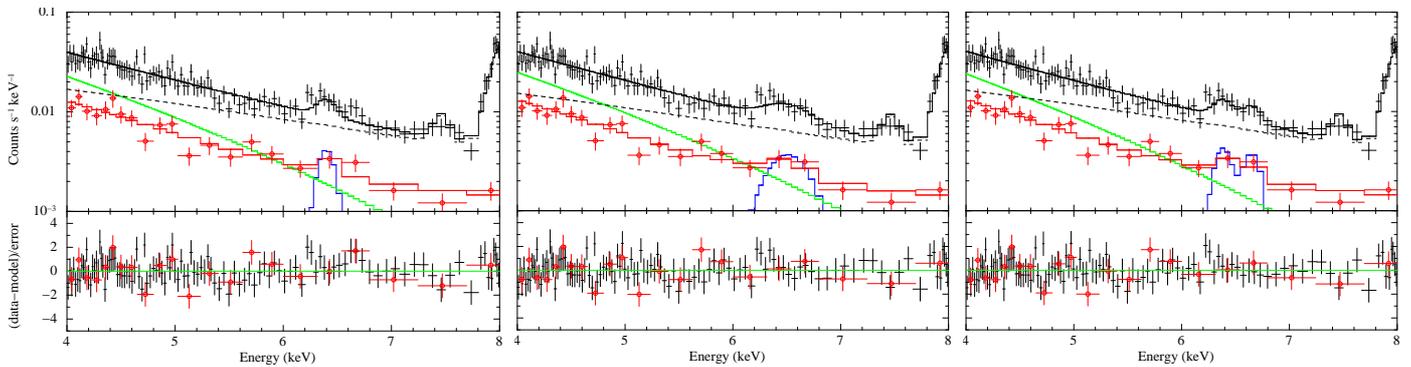

    \centering
\includegraphics[bb=75 10 590 710, clip,
angle=-90,width=0.358\hsize]{final_FeK_shellE_gauss1_bg.ps}
\includegraphics[bb=75 90 590 710,clip,
angle=-90,width=0.316\hsize]{final_FeK_shellE_gauss2_bg.ps}
\includegraphics[bb=75 90 590 710,clip,
angle=-90,width=0.316\hsize]{final_FeK_shellE_gauss3_bg.ps}
\caption{Spectrum of \snr\ around 6.4~keV. The bremsstrahlung continuum is 
shown in green, Fe~K emission in blue. Black points and line are the pn data and 
best-fit model, respectively. All other background contributions are shown by 
the dashed black line. As in Fig.\,\ref{fig_background}, we only show one 
example for MOS (red). Residuals are displayed in the bottom panels. Models used 
for the iron emission (from left to right): Gaussian with width fixed to zero, 
Gaussian with free width, two Gaussians with width fixed to zero. See best-fit 
parameters in Table~\ref{table_FeK}.}
\label{fig_FeK}
\end{figure*}

We start the analysis with the ``Shell E'' spectrum, which has the best signal. 
With a width fixed to zero\,\footnote{The only broadening is from the finite 
spectral resolution of $\approx 150$~eV.}, the best-fit Gaussian has a centroid 
energy of 6405$_{-49} ^{+52}$~eV (Table~\ref{table_FeK}). However, this model 
leaves residuals around the centroid energy, as shown in Fig.\,\ref{fig_FeK} 
(left). When the width ($\sigma$) of the line is freed, the fit improves 
significantly ($\Delta$C$= -12$) for $\sigma = 191_{-\ 72} ^{+131}$~eV, with a 
centroid energy shifted to 6519$_{-79} ^{+81}$~eV (Fig.\,\ref{fig_FeK}, middle). 
The data can be fit equally well or slightly better ($\Delta$C$= -5.3$) by using 
two zero-width Gaussians. In that case, one is at low energy ($\sim6.4$~keV) and 
the second is close to 6.7~keV (Fig.\,\ref{fig_FeK}, right). In the south-west 
region, only a narrow high-energy component is detected; there is no improvement 
when thawing the line width (Table~\ref{table_FeK}). We note that in the 
population study of \citetalias{2014ApJ...785L..27Y}, the Gaussian width was 
left free in all the fits (Hiroya Yamaguchi, personal communication).

\begin{figure}[t]
    \centering
\includegraphics[bb=75 10 590 705, clip, angle=-90, 
width=0.995\hsize]{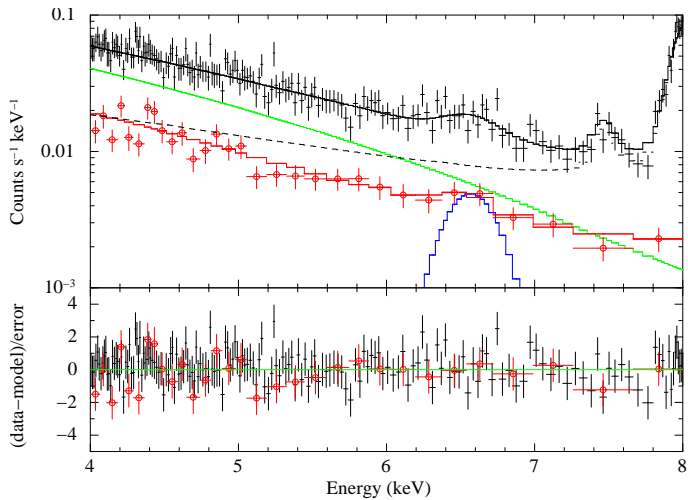}
\caption{Same as Fig.\,\ref{fig_FeK} for the ``Bright'' region. The model used 
for the iron emission is a Gaussian with free width.}
\label{fig_FeK2}
\end{figure}

We also extracted a spectrum covering \emph{both} the east and south-west 
regions (``Bright'' region in Table~\ref{table_FeK}), to measure the integrated 
properties of the Fe~K emission without adding too much background from the 
interior and ``faint NW'' regions where the line is not detected. We found a 
best-fit centroid at 6577$_{-70} ^{+73}$~eV and width of 155$_{-\ 57} 
^{+103}$~eV (Fig.\,\ref{fig_FeK2}). In this region, a good fit is also obtained 
with the two-Gaussian model, for centroids of 6411~$\pm 46$~eV and 
6670$_{-41} ^{+34}$~eV and about the same flux for both components. The total 
flux in the line is $(6.1_{-1.9} ^{+1.8}) \times 10^{-6}$ 
photon~cm$^{-2}$~s$^{-1}$\revisiontwo{, less than 8\,\% of which could be 
attributed to a Galactic ridge contamination}. Most of the flux thus originates 
from the eastern shell. The upper limit on the flux for a 6.6~keV line in the 
remaining regions with no detection is $2 \times 10^{-6}$ 
photon~cm$^{-2}$~s$^{-1}$.

\revision{We attempted to pinpoint the spatial origin of the low-energy 
component (6.4~keV) of the line in the ``Shell E''. The limited statistics and 
spatial resolution prevent us from creating narrow band images above 6~keV. 
Instead, we split the ``Shell E'' region into an inner and outer arcs of same 
width ($\sim 1$\arcmin), and repeated the analysis with the models 
described above. A zero-width Gaussian fit results in a higher ionisation in 
the inner region ($6615_{-45} ^{+28}$~eV) compared to the outer one ($6395_{-45} 
^{+26}$~eV), but the spread reduces somewhat when using the model with free 
width that is slightly preferred. With a two-Gaussian fits, the $\sim 6.65$~keV 
component has a higher flux than the 6.4~keV one in the inner region, while the 
opposite holds in the outer region. This suggests a higher ionisation in the 
inner region, but we stress that the evidence for this are rather marginal due 
to the poor statistics.}
We discuss the complex nature of the Fe~K emission in relation with other SNRs 
in Sect.\,\ref{discussion_FeK}.

\begin{table}[t]
\caption{Fe K line properties.}
\label{table_FeK}
\centering
\begin{tabular}{l c c c}
\hline\hline
\noalign{\smallskip}
 & Shell E & Shell SW & Bright\\
\noalign{\smallskip}
\hline
\noalign{\smallskip}
\multicolumn{4}{c}{Width fixed to zero} \\
\noalign{\smallskip}
\hline
\noalign{\smallskip}
$E$ (eV) &  6405$_{-49} ^{+52}$ & 6690$_{-85} ^{+109}$ & 6645$\pm 43$\\
Norm\tablefootmark{a} & 2.37$_{-0.89} ^{+0.94}$ & 1.00$_{-0.67} ^{+0.79}$ & 
3.38$_{-1.19} ^{+1.26}$ \\
C / d.o.f. & 4039.6/3576 & 3868.5/3573 & 3979.1/3573\\
\noalign{\smallskip}
\hline
\noalign{\smallskip}
\multicolumn{4}{c}{Free width} \\
\noalign{\smallskip}
\hline
\noalign{\smallskip}
$E$ (eV) & 6519$_{-79} ^{+81}$ & 6690$_{-82} ^{+114}$ & 6577$_{-70} ^{+73}$  \\
$\sigma$ (eV) & 191$_{-72} ^{+131}$ & 0 $(< 243)$ & 155$_{-57} ^{+103}$ \\
Norm & 5.29$_{-1.89} ^{+2.56}$ & 1.01$_{-0.67} ^{+0.79}$ & 6.11$_{-1.92} 
^{+1.79}$\\
C / d.o.f. & 4027.6/3575 & 3868.4/3572 & 3970.9/3572\\
\noalign{\smallskip}
\hline
\noalign{\smallskip}
\multicolumn{4}{c}{Two zero-width Gaussians} \\
\noalign{\smallskip}
\hline
\noalign{\smallskip}
$E_1$ (eV) & 6386$_{-46} ^{+36}$ & --- & 6411$\pm 46$\\
Norm$_1$  & 2.38$_{-0.87} ^{+0.94}$ & --- & 2.52$\pm 1.15$\\
$E_2$ (eV) & 6660$_{-46} ^{+43}$ & --- & 6670$_{-41} ^{+34}$ \\
Norm$_2$  & 2.32$_{-0.89} ^{+0.98}$ & --- & 3.32$\pm 1.21$\\
C / d.o.f. & 4022.3/3574 & --- & 3964.8/3571\\
\noalign{\smallskip}
\hline
\end{tabular}
\tablefoot{
\tablefoottext{a}{Normalisation of the line in units of 
$10^{-6}$~photon~cm$^{-2}$~s$^{-1}$.}
}
\end{table}

\section{Discussion}
\label{discussion}

\subsection{Type of progenitor}
\label{discussion_progenitor}
\begin{figure}[t]
    \centering
\includegraphics[width=0.995\hsize]{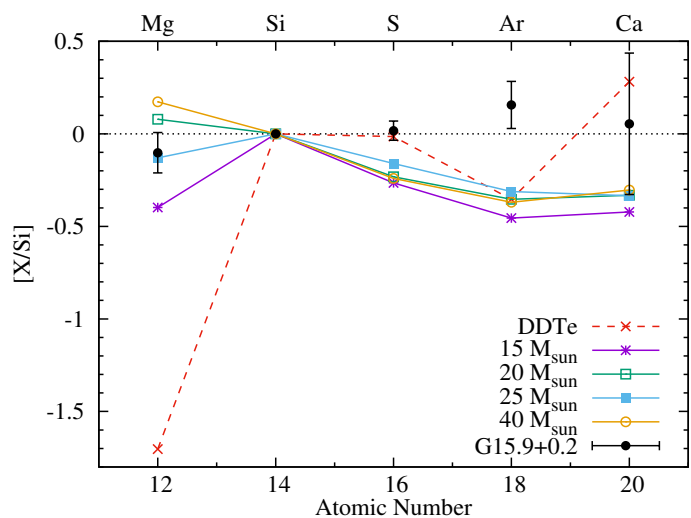}
\caption{Best-fit abundance ratios, relative to Si, measured in the integrated 
spectrum of \snr (black dots). Solid lines show the expected yields of CC SNe 
with various main sequence mass progenitors. The red dashed line is the 
prediction of a type Ia explosion.}
\label{fig_nucleosynthesis}
\end{figure}

The abundance pattern in the ejecta, as revealed by X-ray spectroscopy, contains 
clues to the progenitor type of \snr. In Fig.\,\ref{fig_nucleosynthesis}, we 
plot abundances relative to Si (by number), in the form {[X/Si]} $\equiv \log 
\left[ (\mathrm{X/Si}) / (\mathrm{X/Si})_{\odot} \right]$, using the results of 
the fiducial model in the integrated spectrum. We add the abundance pattern 
predicted by nucleosynthesis models of CC and type Ia SNe; yields of CC SNe of 
various progenitor masses are taken from 
\citet{2006NuPhA.777..424N}\,\footnote{Yields Table 2013, available at 
\url{http://star.herts.ac.uk/~chiaki/works/YIELD_CK13.DAT}.}. For type Ia SNe, 
we used abundances resulting from delayed-detonation and pulsed 
delayed-detonation models, called DDTe and PDDe in \citet{2003ApJ...593..358B}; 
only the former is shown in Fig.\,\ref{fig_nucleosynthesis}. Mg/Si ratios were 
not given originally in \citet{2003ApJ...593..358B}, but are taken from 
\citet{2006ApJ...646..982R}.

What best sets apart type Ia and CC SNe yields is the abundance of magnesium 
relative to silicon. All type Ia models produce very little Mg as it is mostly 
incinerated by the thermonuclear burning front. In massive stars, Mg is mainly 
produced in explosive C/Ne burning and hydrostatic (i.e. pre-SN) burning of C 
and Ne\,\footnote{Large amounts of O and Ne are also produced at the same sites, 
but their X-ray emissions are absorbed and absent in \snr.}, so that the Mg 
yield varies strongly with the progenitor mass. The {[Mg/Si]} ratio in \snr 
($-0.1\pm0.4$) is 40 times that obtained in the DDTe model and 600 times that of 
the PDDe model, while consistent with several CC models of 
\citet{2006NuPhA.777..424N}, for instance a progenitor mass of 18~\msun, 
25~\msun, or 30~\msun. One caveat is that the abundance ratios in 
Fig.\,\ref{fig_nucleosynthesis} are those of the \emph{shocked} ejecta, the only 
material we have access to with X-ray observations. If some amount of Mg is 
missing, the true Mg/Si ratio will be higher and even less compatible with a 
type Ia progenitor. Conversely, if Si is underestimated, the Mg/Si will 
decrease. It is, however, very unlikely that a sufficient quantity of unshocked 
Si is present to produce a Mg/Si ratio of more than one order of magnitude below 
our measured value. Therefore, the Mg/Si ratio provides a strong indication that 
the progenitor of \snr was a massive star, strengthening the physical 
association of the central source CXOU~J181852.0$-$150213 with the SNR.

The ratios of S, Ar, and Ca to Si provide less information because these 
elements are products of explosive oxygen and silicon burning with little 
variation over the progenitor mass range. The variation is mainly due to the 
additional Si contributions from hydrostatic burning 
\citep[][]{1996ApJ...460..408T}. Ca/Si and Ar/Si ratios are also affected by the 
metallicity of the progenitors. The pre-explosion neutron excess $\eta$, mostly 
contributed by $^{22}$Ne, increases linearly with the initial metallicity 
\citep{2003ApJ...590L..83T,2008ApJ...680L..33B}. At increasing $\eta$, yields of 
heavy neutron-rich elements (e.g. Mn and Ni) increase, that of Ca and Ar 
decline, while that of Si remains constant 
\citep{2014ApJ...787..149D,2016ApJ...824...59M}. The high Ar/Si and Ca/Si 
observed in \snr\ could partly be explained by a progenitor with a lower than 
solar metallicity\,\footnote{These arguments come from studies of 
\emph{thermonuclear} SNe. However, the effect of neutronisation on Ca yield 
should qualitatively be the same in the incomplete Si-burning and explosive 
O-burning layers in a CC SN, where most of Ca is produced 
\citep{1996ApJ...460..408T}.}. However, we cannot constrain the progenitor mass 
(or metallicity) very well because of the large uncertainties (particularly 
Ar/Si and Ca/Si). Formally, the model with a progenitor mass of 25~\msun\ 
provides a good match for magnesium and predicts the highest ratios of S, Ar, 
andCa to Si, so it is marginally favoured.

\subsection{Evolution of Fe~K lines in SNRs}
\label{discussion_FeK}
In a study of the Fe~K emission in SNRs with \textit{Suzaku},
\citetalias{2014ApJ...785L..27Y} showed that the centroid energy separates the 
two types of SNRs; the Fe-rich ejecta in Type Ia remnants are significantly less 
ionised than in core-collapse SNRs. This can naturally be explained if the 
ambient medium around CC SNRs is modified (namely made denser) by mass loss 
episodes from the progenitors, while relatively unaffected around type Ia SNe, 
because the ionisation timescale is controlled by $n_e$.

\begin{figure}[t]
    \centering
    \includegraphics[bb=20 110 470 622, clip, 
width=0.995\hsize]{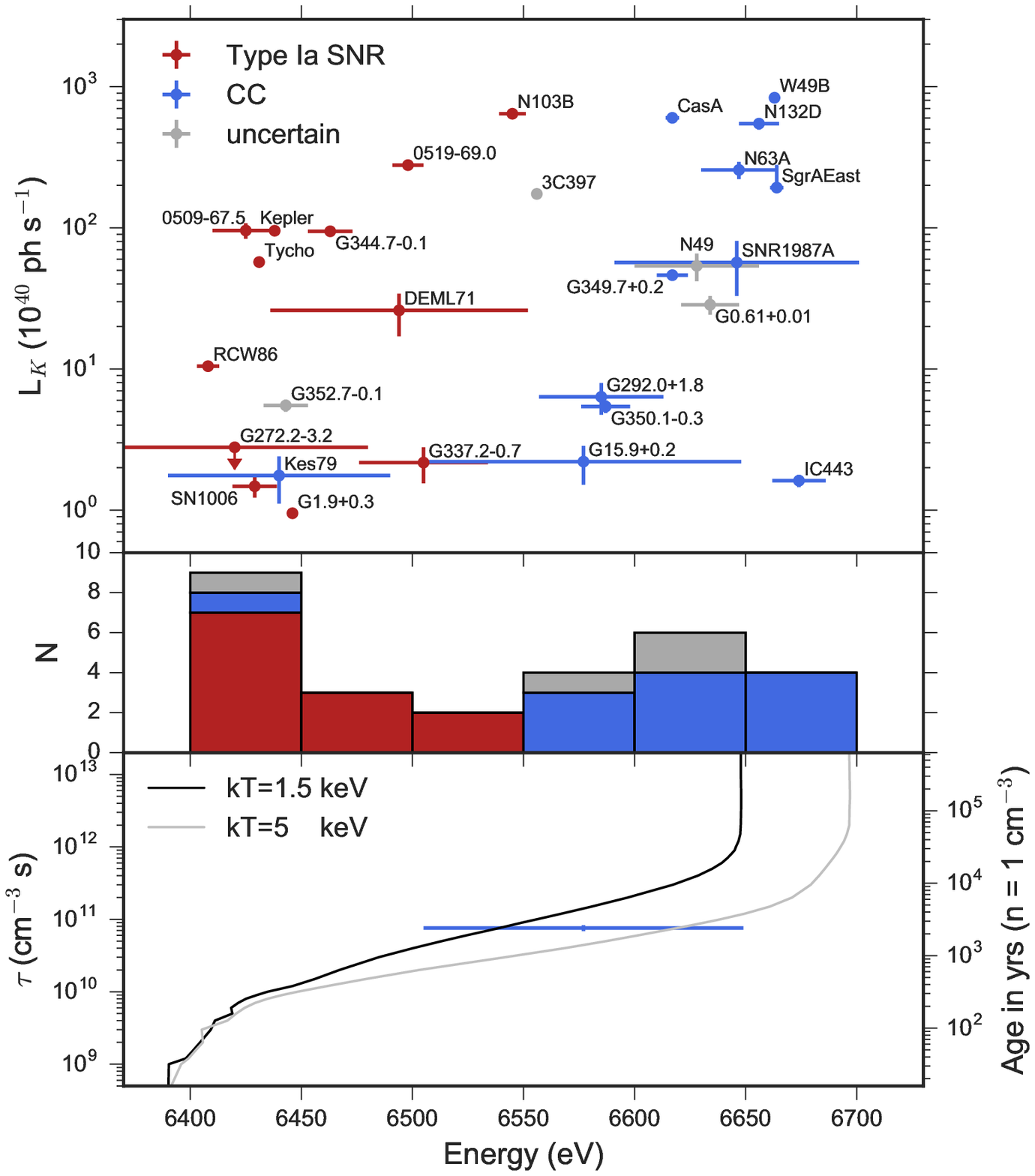}
\caption{
\emph{Top panel\,:} Centroid energy and luminosity of Fe~K lines detected in 
SNRs. The main sample and values obtained with \emph{Suzaku} are taken from 
\citet{2014ApJ...785L..27Y}, augmented with \xmm results for DEM~L71 
\citep{2016A&A...585A.162M} and \snr (this work)\revision{, and \emph{Suzaku} 
observations of Kes~79 \citep{2016PASJ...68S...8S} and \object{SNR 
G272.2$-$3.2} \citep{2016PASJ...68S...7K}. For the latter, only an upper limit 
is shown since the distance is poorly constrained
\citep[2 to 10~kpc, e.g.][]{2001ApJ...552..614H}}. Objects in red and blue are 
type Ia and CC SNRs, respectively. Those in grey are uncertain. 
\emph{Middle panel\,:} Histogram showing the distribution of Fe~K centroid 
energy in SNRs using the same colour code as above.
\emph{Bottom panel\,:} Expected centroid energy of the Fe~K line in a 
\texttt{vpshock} model as function of $\tau$, for $kT = 1.5$~keV (black) and 
5~keV (grey). The best-fit $\tau$ and range of centroid energy of \snr\ are 
indicated. The right-hand $y-$axis translates the ionisation timescale $\tau$ 
into an age, for a uniform ambient density of $n_e = 1$~cm$^{-3}$.
}
\label{fig_FeKsample}
\end{figure}

Since there is strong evidence, as shown above, that \snr originates in a 
core-collapse SN, a high centroid energy is expected for the Fe~K line. Instead 
we found an intermediate value of 6577$_{-70} ^{+73}$~eV when following the same 
method as \citeauthor{2014ApJ...785L..27Y} (i.e. one free-width Gaussian on the 
integrated SNR spectrum), making \snr\ the CC SNR with the lowest Fe~K centroid 
\revision{among the sample of \citetalias{2014ApJ...785L..27Y}} 
(Fig.\,\ref{fig_FeKsample}, top panel). The moderate spatial resolution enabled 
by the existing \xmm\ data reveals that the Fe~K emission is non-uniform: In the 
``Shell E'' region, we detect both $\sim$6.4~keV and $\sim$6.65~keV signals 
attributable to contributions from pooly and highly ionised iron, which in the 
overall spectrum produce a line at the observed centroid of 6.58~keV.

\begin{figure*}[t]
  \begin{center}
    \includegraphics[width=0.69\hsize,bb = 172 254 772 520, clip]
    {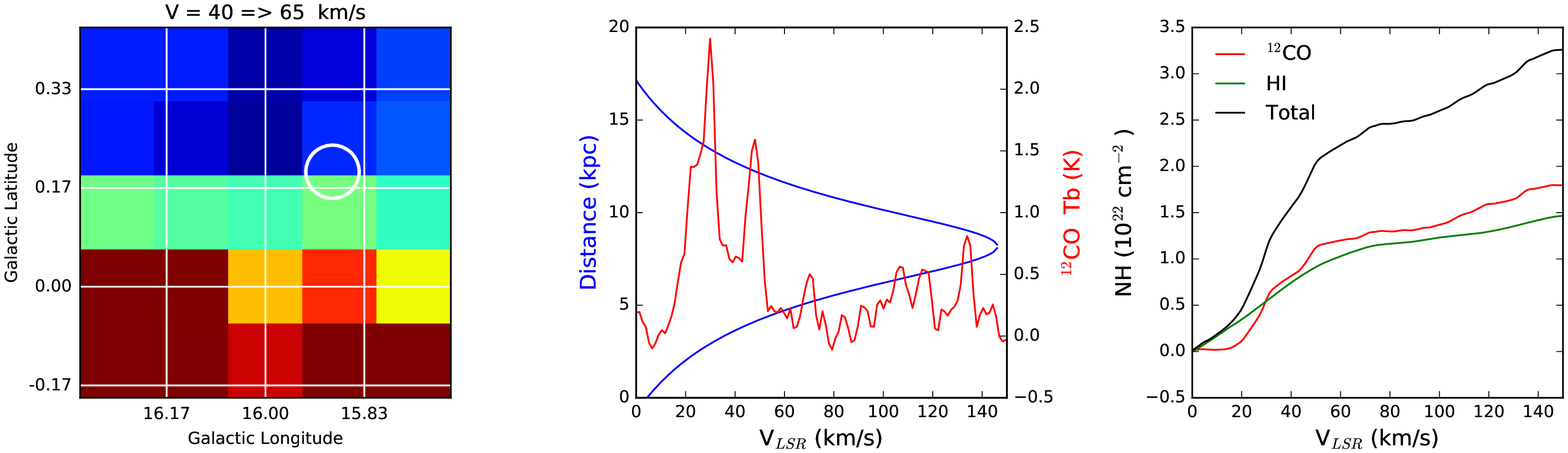}
    \includegraphics[width=0.30\hsize]{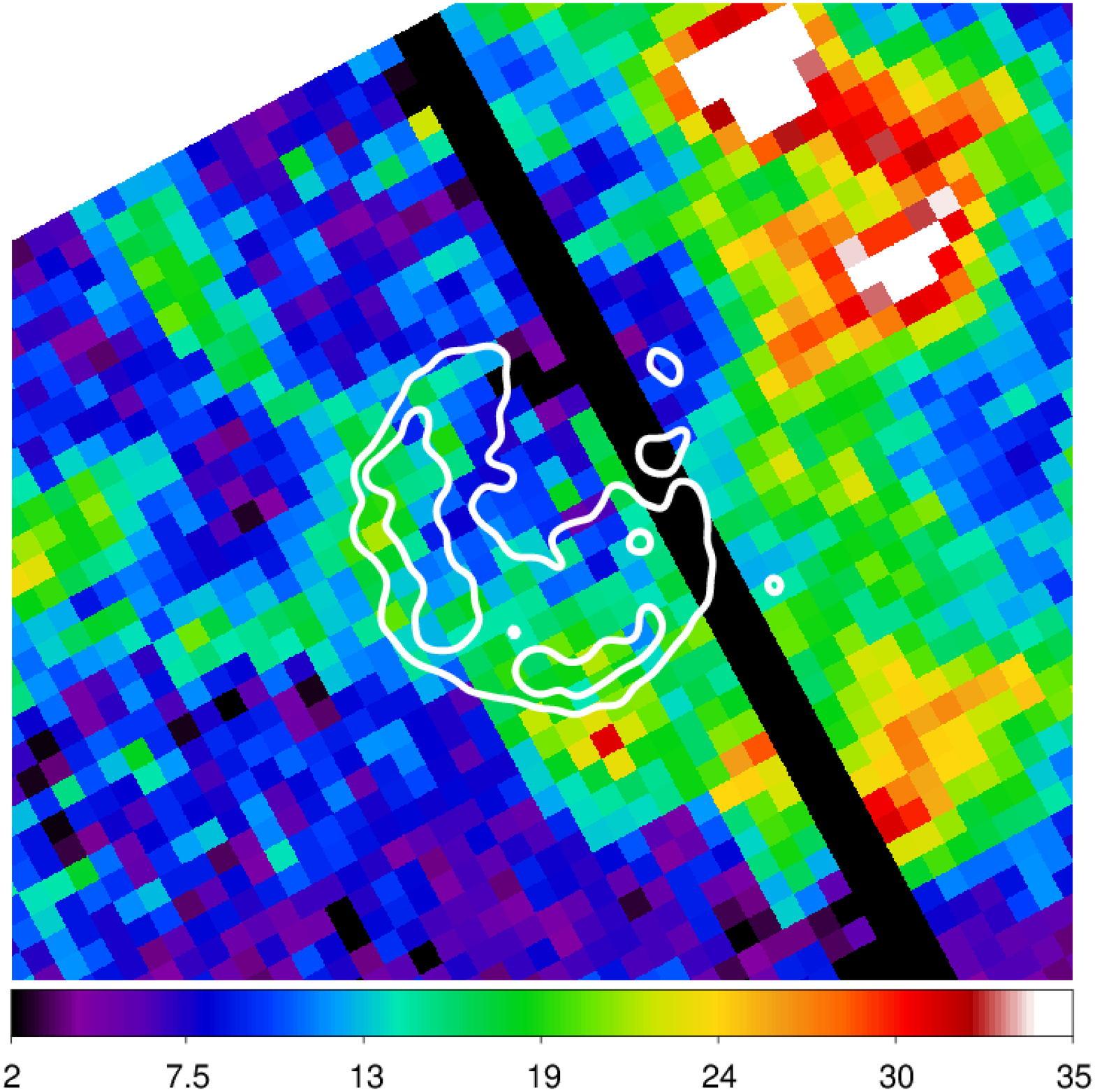}
  \end{center}
  \caption{\emph{Left\,:} $^{12}$CO spectrum at the position of the source 
(red) with corresponding kinematic distance (blue). \emph{Middle\,:} 
\ion{H}{I}, $^{12}$CO, and total column densities towards \snr as a function of 
$V_{\rm LSR}$. \emph{Right\,:}  $^{13}$CO intensity map in the velocity range 
18~km~s$^{-1}$ < $V_{\rm LSR}$ < 32~km~s$^{-1}$, in units of K\,km\,s$^{-1}$. 
X-ray contours are overlaid in white. The black stripe comes from bad 
columns which have been set to 0.
}
  \label{fig_CO-HI}
\end{figure*}

\revision{The detection of a 6.4~keV line from a CC SNR is remarkable. Such a 
line has been found in CC SNRs \object{Kes~79} \citep{2016PASJ...68S...8S} and 
\object{HESS~J1745$-$303} \citep{2009ApJ...691.1854B}, but it was proposed not 
to be of ionised plasma origin. In Kes~79, the 6.4~keV emission is attributed to 
K-shell ionisation of neutral iron by the interaction of locally accelerated 
protons with a nearby molecular cloud, while \citet{2009ApJ...691.1854B} 
suggested that the line in HESS~J1745$-$303 was from an X-ray reflection nebula 
reflecting X-rays from previous Galactic centre activity off a molecular cloud. 
Neither of these scenarios are likely for \snr: There are no striking spatial 
correlations of the 6.4~keV emission with molecular clouds as in Kes~79 (see 
Fig.\,\ref{fig_CO-HI}), nor reported detection of OH maser to support an
interaction with molecular clouds \citep{1997AJ....114.2058G}. An X-ray 
reflection nebula origin, as in HESS~J1745$-$303, is also not favoured given 
the lack of a suitable nearby hard X-ray source, and the poor spatial 
correlation between the 6.4~keV emission and molecular material.}

\revision{In \snr, an ionised plasma origin is favoured. The variations in 
ionisation between east and west regions, and inner/outer parts of ``Shell E'' 
(Sect.\,\ref{results_spectral_FeK}) points to a range of $n_et$ due to 
variations in the ambient density encountered by the shock, or to different 
times since the plasma was shocked. The material in the east could be part of a 
cavity that the shock encountered recently, akin to the case in the type Ia SNR 
\object{RCW~86} \citep{2014MNRAS.441.3040B}. Such density variations and/or 
cavities} are likely to be common, especially around young CC SNRs, but are hard 
to observe, except in the brightest and more extended sources. As another 
instance, \citet{2005ApJ...618..321S} found different Fe~K centroids in eastern 
and western regions of 3C~397, attributed to density gradients.

\citet{2015ApJ...803..101P} have modelled CC SNe exploding in CSM modified by 
progenitors' winds and followed their dynamics (i.e. shock radius) and spectral 
evolution of Fe K emission (centroid and luminosity). Although these models were 
able to reproduce the higher centroids (above 6550~eV) observed in CC SNRs, 
there were many outliers where either the correct shock radius and centroid were 
not reproduced by the models, or where the Fe K luminosity was severely 
underpredicted. Placing \snr\ in that context, the bulk Fe~K properties (in 
particular its low centroid) and size of about 4~($D$\,/\,5~kpc)~pc (see 
Sect.\,\ref{results_morphology}) can be reproduced by their SN~1987A or 
\object{SN~1993J} models \citep[e.g. Fig.\,2,][]{2015ApJ...803..101P}. The low 
Fe~K luminosity of \snr\ ($\sim 2 \times 10^{40} 
$~($D$\,/\,5~kpc)~photon~s$^{-1}$) can be reproduced by most of their models in 
the lower mass loss rate/higher wind velocity cases.

Regardless of type, the distribution of Fe~K centroids in SNRs is peaked at the 
low- and high-energy ends (see Fig.\,\ref{fig_FeKsample}, middle panel) with 
relatively few objects between 6.5~keV and 6.6~keV. This is due to the rapid 
transition between a roughly 6.4~keV centroid (no iron above the \ion{Fe}{XVII} 
state) and a 6.6$-$6.7~keV centroid (most iron above \ion{Fe}{XX}). To 
illustrate this, we simulated emission from a pure-iron ionising plasma 
(\texttt{vpshock} in XSPEC) at $kT = 1.5$~keV and 5~keV, with $\tau$ between 
10$^8$~s~cm$^{-3}$ and $5 \times 10^{13}$~s~cm$^{-3}$. Fake spectra were 
produced using the EPIC-pn response and analysed in the same fashion as in 
\citetalias{2014ApJ...785L..27Y} and in this work. The curves of the expected 
centroids are shown in the bottom panel of Fig.\,\ref{fig_FeKsample}. The Fe~K 
transition  occurs briefly, or more specifically over a small $\tau$ interval, 
between a few 10$^{10}$~s~cm$^{-3}$ and a few 10$^{11}$~s~cm$^{-3}$. For SNRs 
expanding in a constant ambient density of $n_e = 1$~cm$^{-3}$, this happens at 
an age of $\sim 1000 - 5000$~yr. The oldest type Ia SNRs in the sample of 
\citetalias{2014ApJ...785L..27Y} are from 4500 to 6000~yr old (\object{SNR 
G337.2$-$0.7} and \object{SNR G344.7$-$0.1}). They are associated with 
relatively low density \citep[$n_e \sim$~0.3 to 
0.8~cm$^{-3}$,][]{2006ApJ...646..982R,2011A&A...531A.138G} so that a low Fe 
ionisation is expected even at this age.

The possibility remains open for a type Ia SNR in a denser medium ($\gtrsim 
2$~cm$^{-3}$) to reach a high ionisation of iron within a few thousand years and 
populate the Fe~K centroid region dominated so far by CC SNRs. Without 
definitive evidence as to its type \citep{2012ApJ...748..117P}, \object{N49} 
might be such an object, \revision{since it is interacting with a relatively 
high density \citep[at least 2--25~cm$^{-3}$, 
][]{2004AJ....128.1615S,2016ApJ...826..150D}}. Symmetrically, it can not 
excluded that some CC SNRs are found at the low ionisation end. Part of \snr, 
for instance, still has emission at 6.4~keV, while \object{SNR~G352.7$-$0.1} 
has an Fe~K centroid at 6.44~keV and is also of uncertain type; some studies 
suggest that it is the remnant a CC SNR 
\citep{2009A&A...507..841G,2014ApJ...782..102P}. Another caveat is that the 
current sample of Fe~K emitting objects inherently contains some age bias, the 
youngest objects being preeminently of type Ia. There are six type Ia SNRs 
younger than $\sim 1000$~yr, but only three CC SNRs below the 2000~yr mark. 
Together with the lower density expected around type Ia SN progenitors, 
this bias contributes to separate the Fe~K centroids of both classes. We 
conclude that the Fe~K centroid-luminosity diagram provides valuable information 
about an SNR, but care should be taken when using it as a typing tool without 
any knowledge of the surrounding medium or indication of age. This is 
particularly true for unresolved objects, e.g. young SNRs in external galaxies, 
\revision{or objects where the nature of the Fe~K emission is unknown and might 
not be from an ionised plasma, as is the case in Kes~79 and HESS~J1745$-$303}.


\subsection{Constraining the age of and distance to \snr}
\label{discussion_age_distance}
To constrain the distance to the SNR, we compared the absorption along the line 
of sight derived from the X-ray and from \ion{H}{I} and $^{12}$CO (1-0) 
observations. Figure~\ref{fig_CO-HI} (left panel) shows the velocity spectra at 
the SNR centre (RA = 18\hour\,18\minute\,53.8\second, DEC = 
$-$15\degr\,01\minute\,38\second) integrated within a radius of 3\arcmin. The 
$^{12}$CO observations are from the CfA survey \citep{2001ApJ...547..792D} and 
the \ion{H}{I} data from the SGPS survey \citep{2006ApJS..167..230H}. The 
corresponding kinematic distances assuming the circular Galactic rotation model 
of \citet{2009A&A...499..473H} with a distance to the Galactic center of 8~kpc 
are also shown. We note that the bulk of the material along the line of sight 
is at radial velocities relative to the \textit{local standard of rest} (LSR) of 
$V_{\rm LSR}$ < 60~km~s$^{-1}$ which is equivalent to a lower limit on the 
distance of 5~kpc. The cumulative $N_{\rm H}$ derived from \ion{H}{I} and 
$^{12}$CO as a function of $V_{\rm LSR}$ is shown in Fig.\,\ref{fig_CO-HI} 
(middle panel) where all the material is assumed to be at the near distance 
allowed by the Galactic rotation curve, providing therefore a lower limit on the 
distance. The CO-to-H$_{2}$ mass conversion factor and the \ion{H}{I} brightness 
temperature to column density used is respectively of 
$1.8\times10^{20}$~cm$^{-2}$~K$^{-1}$~km$^{-1}$~s \citep{2001ApJ...547..792D} 
and $1.82\times10^{18}$~cm$^{-2}$~K$^{-1}$~km$^{-1}$~s 
\citep{1990ARA&A..28..215D}.

Most of the absorption along the line of sight is due to the structures at 
$V_{\rm LSR}$ < 60~km s$^{-1}$, representing an integrated $N_{\rm 
H}=2.2\times10^{22}$~cm$^{-2}$. With an X-ray absorption value of 
$4.58\pm0.11\times10^{22}$~cm$^{-2}$ (see Table~\ref{table_spectral_results}), 
we argue that the SNR is behind these structures and set a strict lower limit to 
its distance at 5~kpc. \revision{This is consistent with the results of a carbon 
atmosphere fit to the CCO emission \citep{2016A&A...592L..12K} that yields a 
large distance (although highly uncertain) of 10$^{+9}_{-5}$~kpc.}

However, the total $N_{\rm H}$ integrated along the entire velocity range 
derived from \ion{H}{I} and $^{12}$CO amounts to $\sim3.5\times10^{22}$ 
cm$^{-2}$, still significantly less than the X-ray value. This discrepancy could 
be due to an inadequate value of the CO-to-H$_{2}$ mass conversion factor which 
can vary within the Galaxy \citep[e.g.][]{2015MmSAI..86..616R} or missing 
material not traced by HI or $^{12}$CO emission. In particular in dense 
molecular clouds, the core of the cloud is not well traced by $^{12}$CO owing 
to saturation of the emission.

The $^{13}$CO (1-0) line emission provides a tracer to investigate the core of 
the dense molecular clouds. In Fig.\,\ref{fig_CO-HI} (right panel), we show the 
high-resolution $^{13}$CO intensity map\footnote{0.4\arcmin\ pixel size for the 
$^{13}$CO map vs. 7\arcmin\ pixels for the $^{12}$CO map.} from the GRS survey 
\citep{2006ApJS..163..145J}. This map was obtained by integrating velocities 
between 18 km~s$^{-1}$ < $V_{\rm LSR}$ < 32~km~s$^{-1}$ corresponding to main 
peak of $^{13}$CO emission located at the same velocity as the brightest 
$^{12}$CO peak shown in Fig.\,\ref{fig_CO-HI}. This material is likely in the 
foreground of the SNR. Small clouds are seen at the position of the shell E and 
shell SW regions, explaining the spatial variations of $N_H$ found in 
Sect.\,\ref{results_spectral_fits}.

To investigate the age of the SNR in more detail, we used the equations
describing the evolution of the forward shock $R$ presented by
\citet{1999ApJS..120..299T} and generalised  to a steady stellar wind
environment with $\rho(r) \propto r^{-2}$ by \citet{2003ApJ...597..347L} \&
\citet{2012ApJ...746..130H}. We fixed the explosion energy to $E=10^{51}$ ergs,
$n=9$, and $s=2$, where $n$, and $s$ are the ejecta and ambient medium density
profile ($\rho(r) \propto r^{-n,s}$). The ejecta mass was fixed  to 14
$M_{\sun}$ (for an initial progenitor mass of 25  $M_{\sun}$) according to mass
budget models of core-collapse supernovae by \citet{2016ApJ...821...38S}.

The ambient medium density is derived from the X-ray thermal emission of the
``faint NW'', the least affected by ejecta and therefore best representing
shocked ISM. The normalisation of the fiducial model in that region
(Table~\ref{table_spectral_results}) is proportional to the volume emission
measure $\int n_e n_H dV$, which can be linked to the pre-shock ambient hydrogen
density $n_{\mathrm{H,}0}$ assuming a Sedov profile \citep[e.g.][Eq.
1]{2014A&A...561A..76M}. With the uncertainties on the normalisation in the
``faint NW'' region, we measure a range of $n_{\mathrm{H,}0} = (0.36 - 0.70)$
($D$\,/\,5~kpc)$^{-1/2}$~cm$^{-3}$.

Considering the lower limit on the distance of 5 kpc derived above, we obtain a
lower limit on the age of 2~kyr. This reproduces the observed angular 
distance (3.7\arcmin) between the CCO  and the tip of the faint NW region 
(Fig.\,\ref{fig_xray_extraction}). If the SNR is near the Galactic centre, at a 
distance of 8.5~kpc as assumed by \citetalias{2006ApJ...652L..45R}, the age 
estimate is $\sim$ (4.5--6) kyr. In the case where the shock has evolved only a
small fraction of its lifetime in the steep wind profile, and quickly after
evolved in the uniform shocked wind region (assuming therefore $s = 0$), the
corresponding ages are $t_{SNR} >$ 1.6~kyr and $t \sim (3.5 - 4)$~kyr for a 
distance of 5 and 8.5~kpc, respectively.

\section{Summary}
\label{summary}
We have studied the Galactic SNR G15.9+0.2, observed serendipitously with \xmm. 
The remnant exhibits a shell morphology with the brightest regions in the east 
and south-west, and the faintest and softest to the north-west. Elemental 
abundances, particularly in the bright regions, are markedly supersolar, 
betraying a SN ejecta origin. Analysis of the ejecta abundance pattern 
establishes SNR G15.9+0.2 as a core-collapse SNR and suggests a progenitor mass 
of 20--25~\msun. This makes the physical association with the CCO 
CXOU~J181852.0$-$150213 detected in its interior very likely.

We report for the first time detection of Fe K emission from SNR G15.9+0.2,
using a careful treatment of all background components to ensure that this
signal is genuinely from the SNR. We detect signal at both the low (6.4~keV) and
high (6.67~keV) ionisation end of the line, with spatial variation across the
SNR. Because SNR G15.9+0.2 is the core-collapse SNR with the lowest overall 
Fe~K centroid energy, caution should be used when typing SNRs based on this
criterion alone.

Matching the X-ray $N_H$ to atomic and molecular gas structures along the
line of sight, we set a conservative limit of 5~kpc for the distance towards the
source. At such distances and for a reasonable set of parameters affecting the
remnant's dynamics, we constrain its age to be likely more than 2000~yr.

\begin{acknowledgements}
The authors thank the anonymous referee for helping us to improve the discussion 
of our results. We thank Carles Badenes for helpful discussions about SN 
nucleosynthesis.
P.\,M. acknowledges support by the Centre National d'\'Etudes Spatiales (CNES). 
This research has made use of Aladin, SIMBAD, and VizieR, operated at the CDS,
Strasbourg, France.
\end{acknowledgements}




\end{document}